\documentclass[12pt]{iopart}
\usepackage{amssymb}

\newcommand{\btr}{\bigtriangledown}

\newcommand{\la}{\lambda}
\newcommand{\lb}{\label}

\newcommand{\al}{\alpha}
\newcommand{\bt}{\beta}
\newcommand{\ka}{\kappa}
\newcommand{\pa}{\partial}
\newcommand{\ga}{\gamma}
\newcommand{\si}{\sigma}

\newcommand{\fr}{\frac}

\newcommand{\f}{{f}}
\newcommand{\bw}{\begin{widetext}}
\newcommand{\ew}{\end{widetext}}
\newcommand{\be}{\begin{equation}}
\newcommand{\ee}{\end{equation}}
\newcommand{\bee}{\begin{equation*}}
\newcommand{\eee}{\end{equation*}}
\newcommand{\ba}{\begin{eqnarray}}
\newcommand{\ea}{\end{eqnarray}}
\newcommand{\bal}{\begin{align}}
\newcommand{\eal}{\end{align}}
\newcommand{\bml}{\begin{multline}}
\newcommand{\eml}{\end{multline}}
\newcommand{\non}{\nonumber}
\newcommand{\va}{\varepsilon}
\newcommand{\ep}{\epsilon}

\newcommand{\cP}{\cal P}

\def\e{{\rm e}}
\begin{document}
\title[Dilaton and axion  bremsstrahlung from collisions of cosmic
(super)strings] {Dilaton and axion  bremsstrahlung from collisions
of cosmic (super)strings}

\author{E Yu Melkumova$^1$, D V Gal'tsov$^2$ and K Salehi$^3$}

\address{Department of Physics, Moscow State University,119899,
Moscow, Russia\footnote{Talk given at 2nd International Conference
On Quantum Theories And Renormalization Group In Gravity And
Cosmology (IRGAC 2006), 11-15 Jul 2006, Barcelona, Spain}}
\ead{$^1$elenamelk@srd.sinp.msu.ru ; $^2$galtsov@phys.msu.ru;
$^3$ksalehi2002@yahoo.com}

\begin{abstract}
We calculate dilaton and axion radiation generated in the
collision of two straight initially unexcited strings and give a
rough cosmological estimate of dilaton and axion densities
produced via this mechanism in the early universe.

\end{abstract}

\section{Introduction}

Recently the early universe models involving strings and branes
moving in higher-dimensional space-times received a renewed
attention \cite{DuKu05}-\cite{GaM01}. In particular, the problem
of the dimensionality of space-time can be explored within the
brane gas scenario \cite{DuKu05}-\cite{Po04}. Another new
suggestion is the possibility of cosmic superstrings with lower
tension than those in the field-theoretical GUT strings
\cite{Po04}. Superstrings as cosmic strings candidates  stimulate
reconsideration of the cosmic string evolution with account for
new features  such as existence of the dilaton and antisymmetric
form fields and extra dimensions. The main  role in this evolution
is played by radiation processes. The radiation mechanism which
has been mostly studied in the past consists in formation of the
excited closed loops  which subsequently loose their excitation
energy emitting gravitons \cite{VaVi85} axions \cite{DaQu89} and
dilatons \cite{Da85}-\cite{ViVa85}.

In this paper we consider the bremsstrahlung mechanism of string
radiation  \cite{GaGrLe93} which works for initially unexcited
strings undergoing a collision. This effect is  similar to
bremsstrahlung under collision of point charges in
electrodynamics. In  the perturbation  expansion in terms of the
fine structure constant, bremsstrahlung is the second order
process. In the case of strings we develop a classical
perturbation scheme for two endless unexcited long strings which
move one with respect to another in two parallel planes being
inclined at an angle. It was shown earlier that in four space-time
dimensions there is no gravitational bremsstrahlung under
collision of straight strings \cite{GaGrLe93}. This can be traced
to absence of gravitons in 1+2 gravity. It is not a coincidence
that in four dimensions there is no gravitational renormalization
of the string tension either \cite{BuDa98}. But there is no such
dimensional argument in the case of the axion field there  such
dimensional argument and it was demonstrated that string
bremsstrahlung takes place indeed \cite{GMK04} within the model in
flat space. Here we extend this result to the full gravitating
case including also the dilaton field. Strings interacts via the
dilaton, axion and graviton exchange. Radiation arises in the
second order approximation in the coupling constants provided the
(projected) intersection point moves with superluminal velocity.
Thus, the string bremsstrahlung can be viewed as manifestation of
the Cherenkov effect.
\section{ String interactions}
Consider a pair of relativistic strings
$$x^{\mu}=x_n^{\mu}(\sigma^a_n),\;\; \mu=0,1,2,3,\;\; \sigma_a=(\tau,
\sigma),\;\; a=0,1,$$ where  $n=1,2$ is the  index labelling the
two strings. The 4-dimensional space-time metric signature $+,---$
and $(+,-)$ for the string world-sheets metric signature. Strings
interact via the gravitational $ g_{\mu\nu} \equiv \eta_{\mu\nu} +
h_{\mu\nu} $, dilatonic $ \phi (x) $ and axion (Kalb-Ramond) field
$B_{\mu\nu}(x)$: \ba \fl S =-\sum_n\int\left\{
 \fr{\mu}{2} \partial_a x^\mu_n \partial_b x^\nu_n
 g _{\mu\nu}\gamma^{ab}\sqrt{-\gamma}\e^{2\al_n\phi}+2\pi\f
  \partial_a x^\mu_n \partial_b x^\nu_n \epsilon^{ab}
  B_{\mu\nu}\right\}d^2 \sigma \non\\+
\int\limits\left\{ 2\partial_\mu\phi\partial_\nu\phi
g^{\mu\nu}+\fr{1}{6}H_{ \mu \nu \rho} H^{ \mu \nu
\rho}\e^{-4\al\phi} - \frac{R}{16\pi G } \right\}\sqrt{-g} \,
d^4x.\label{fgac} \ea
 Here $\mu_n$ are the (bare) string tension
parameters, $\al$ and $f$ are the corresponding coupling
parameters, $\epsilon^{01}=1,\; \ga_{ab}$ is the induced metric on
the world-sheets. In what follows, we linearize the dilaton
exponent as $ \e^{2\al\phi}\simeq 1+2\al\phi $.
\par
The totally antisymmetric axion field strength is defined as \be
H_{ \mu \nu \la}=\partial_\mu B_{\nu\la}+\partial_\nu
B_{\la\mu}+\partial_\la B_{\mu\nu}.\ee Variation of the action
(\ref{fgac}) over $ x^\mu_n $ leads to the  equations of motion
for strings \ba\fl \partial_a\left(\mu
\partial_b x^\nu_n
g_{\mu\nu}\ga^{ab}\sqrt{-\ga}\e^{2\al\phi}+4\pi f
\partial_b x^\nu_n\epsilon^{ab}B_{\mu\nu}\right)\non\\-
\mu\al \partial_a x^\al_n
\partial_b x^\beta_n
g_{\al\beta}\ga^{ab}\sqrt{-\ga}
\e^{2\al\phi}\partial_{\mu}\phi-\fr\mu2 \partial_a x^\al_n
\partial_b x^\beta_n\ga^{ab}\sqrt{-\ga}
\e^{2\al\phi}\partial_\mu g_{\al\beta}=0.\label{steq}\ea Variation
with respect to field variables $ \phi $, $ B_{\mu\nu} $ and $
g_{\mu\nu} $ leads to the dilaton equation: \be\fl
\partial_\mu\left(g^{\mu\nu}\partial_\nu
\phi\sqrt{-g}\right)+\fr{\al}6
H^2\e^{-4\al\phi}+\fr{\mu\al}{4}\int
\partial_a x^\mu_n
\partial_b x^\mu_n
g_{\mu\nu}\ga^{ab}\e^{2\al\phi}\delta^4(x-x_n(\si_n))d^2\sigma=0,\label{fep}\ee
the axion  equation: \be\label{feB}
\partial_\mu\left(H^{\mu\nu\la}\e^{-4\al\phi}\sqrt{-g}\right)+2\pi
f \int \partial_a x^\nu_n \partial_b x^\la_n
\epsilon^{ab}\delta^4(x-x_n(\si_n))d^2\sigma=0\ee and the Einstein
equations  \ba\label{gre} R_{\mu\nu}-\fr12g_{\mu\nu}R=8\pi
G(\stackrel{\phi}{T}{\!}_{\mu\nu}+\stackrel{B}{T}{\!}_{\mu\nu}+
\stackrel{st}{T}{\!}_{\mu\nu}),\ea \bee
\stackrel{st}{T}{\!}_{\mu\nu}=\sum \mu\int
\partial_a x_{\mu \, n}
\partial_b x_{\nu \,
n}\gamma^{ab}\sqrt{-\gamma}\e^{2\al\phi}
\fr{\delta^4(x-x_n(\si_n))}{\sqrt{-g}}d^2\sigma,\eee \ba\fl
\stackrel{\phi}{T}{\!}_{\mu\nu}\!=\!4\left(\partial_\mu\phi\partial_\nu\phi-\fr12
g_{\mu\nu}(\btr\phi)^2\right),\qquad
\stackrel{B}{T}{\!}_{\mu\nu}=\left(H_{\mu\al\bt}H_\nu^{\al\bt}-
\fr16H^2g_{\mu\nu}\right)\e^{-4\al\phi}.\lb{TTT}\ea The constraint
equations for each string read : \ba (\partial_a x^\mu
\partial_b x^\nu - \fr12 \ga_{ab}\ga^{cd}
\partial_c x^\mu \partial_d x^\nu )g_{\mu\nu}=0.\label{con}\ea
\par
Our calculation follows the approach of
\cite{GaGrLe93},\cite{GMK04} and  consists in constructing
solutions of the string equations of motion and dilaton, axion and
graviton iteratively using the coupling constants $\alpha,\,f,\,G$
as expansion parameters.
\par
Denote as  $\stackrel{0}{x}{\!\!}^\mu$ the embedding function of
the non-exited straight string stretched along the direction
$\Sigma^\mu $ and moving as a whole with the four-velocity $
u^{\mu}$. This is the linear function of $ \tau,\,\sigma $:
   \be\label{xo}
   \stackrel{0}{x}{\!\!}^\mu \ = \ d^\mu \ + \ u^\mu \tau \
+ \Sigma^\mu \sigma,  \ee where the constant vector $ d^\mu $ can
be regarded as an impact parameter. The  zero-order space-time
metric is assumed flat $ \stackrel{0}{g}{\!\!}_{\mu\nu} =
\eta_{\mu\nu} $, and the zero-order world-sheet induced metric can
be also made Minkowskian. Indeed, assuming $ \eta_{\mu\nu} x^\mu_a
x^{\nu b}=\delta^a_b $, i.e.
$\eta_{\mu\nu}\Sigma^\mu\Sigma^\nu=-1, \quad \eta_{\mu\nu} u^\mu
u^\nu =1,\quad \eta_{\mu\nu} \Sigma^\mu u^\nu=0, $ one has $
\stackrel{0}{\ga}{\!\!}_{ab}= \pa_a\stackrel{0}{X}{\!\!}^\mu\pa_b
\stackrel{0}{X}{\!\!}^\nu \eta_{\mu\nu}=\eta_{ab}=diag(1,-1)$.
Assuming that zero order (external) dilaton and axion fields are
absent $ \stackrel{0}{\phi}=0,\;\stackrel{0}{B}{\!\!}_{\mu\nu}=0$,
we expand all the field variables $ \phi $, $B_{\mu\nu},\;
h_{\mu\nu}$  starting with the first order: $ \phi=
\stackrel{1}{\phi} + \stackrel{2}{\phi}+ \ldots,\; B_{\mu\nu} =
\stackrel{1}{B}{\!\!}_{\mu\nu} + \stackrel{2}{B}{\!\!}_{\mu\nu}+
\ldots,\; h_{\mu\nu} = \stackrel{1}{h}{\!\!}_{\mu\nu} +
\stackrel{2}{h}{\!\!}_{\mu\nu}+ \ldots $.
\par
The total  dilaton, axion and graviton fields are the sums due to
contributions of two strings: $\phi=\phi_1+\phi_2,\;
B^{\mu\nu}=B^{\mu\nu}_1+B^{\mu\nu}_2,\;
h^{\mu\nu}=h^{\mu\nu}_1+h^{\mu\nu}_2.$ Since in the zero order the
strings are moving freely (\ref{xo}), the first order dilaton $
\stackrel{1}{\phi} _{n} $, axion $\stackrel{1}{B}{\!}^{\mu\nu}_n $
and graviton variables $\stackrel{1}{h}{\!}^{\mu\nu}_n $  do not
contain radiative components. Substituting them into the Eq.
(\ref{steq}) we then obtain the first order deformations of the
world-sheets $\stackrel{1}{x}{\!}^{\mu} $, which are naturally
split into contributions due to dilaton, axion and graviton
exchange: \ba\lb{x1} \stackrel{1}{x}{\!}^{\mu}_n &=&
\stackrel{1}{x}{\!}^{\mu}_{n (\phi)} +
\stackrel{1}{x}{\!}^{\mu}_{n (B)}+\stackrel{1}{x}{\!}^{\mu}_{n
(h)}.\ea The deformation due to the dilaton reads \ba\label{x1p1}
\stackrel{1}{x}{\!}^{\mu}_{n (\phi)}(\tau ,\sigma )=
-i\fr{\al^2\mu }{16\pi^2} \int
\frac{\Delta_{n'}\textit{D}_{n'n}^\mu
\e^{-iq(d_n+u_n\tau+\Sigma_n\sigma)}}{q^2((q
\Sigma_1)^2-(qu_1)^2)} d^4 q , \ea   where $  \Delta_{n'}=
\e^{iqd_{n'}}\delta(q u_{n'})\delta(q\Sigma_{n'}),\;
\textit{D}_{n'n}^\mu= \stackrel{0}{U}_{n'}
  ( q^\mu \stackrel{0}{U}_{n}
+ 2\Sigma^\mu_n(\Sigma_nq) - 2u^ \mu_n(u_nq)), \;
\stackrel{0}U_n=\eta^{\mu\nu}\stackrel{0}U{\!}^{\mu\nu}_{n},\;
\stackrel{0}U{\!}^{\mu \nu}_n = u^\mu_n
u^\nu_n-\Sigma^\mu_n\Sigma^\nu_n,\; \stackrel{0}U_n=2 $. The
corresponding axion contribution  is \be\label{x1B1}
\stackrel{1}{x}{\!}^{\mu}_{n (B)} =- i\fr{2f^2}{ \mu }\int
\frac{X^{\mu}_{n'n}\Delta_{n'}\e^{-iq(d_n+u_n\tau+\Sigma_n\sigma)}
} {q^2[(q\Sigma_1)^2-(qu_1)^2]} d^4q, \ee where $X^{\mu}_{n'n}=
q^\mu A_{n'n} + B_{n'n}\Sigma^\mu_{n'} + C_{n'n}u_{n'}^\mu,\,\,
A_{n'n}=(u_nu_{n'})(\Sigma_n \Sigma_{n'})-( \Sigma_n u_{n'})(u_n
\Sigma_{n'}), $ $B_{n'n}=(qu_n)(u_{n'}
\Sigma_n)-(\Sigma_nq)(u_nu_{n'}), \,\, C_{nn'}=(u_n \Sigma_{n'})(
\Sigma_nq)-(qu_n)(\Sigma_n \Sigma_{n'})$ and the gravitational
contribution is
 \be\label{x1h1} \stackrel{1}{x}{\!}^{\mu}_{n (h)} =- i\fr2\pi
G{ \mu}\int
\frac{Z^{\mu}_{n'n}\Delta_{n'}\e^{-iq(d_n+u_n\tau+\Sigma_n\sigma)}}
{q^2[(q\Sigma_1)^2-(qu_1)^2]} d^4q, \ee where $Z^\mu_{n'n} =(
\stackrel{0}{W}{\!}_{n'}^{ \al\bt}q^\mu
-2\stackrel{0}{W}{\!}_{n'}^{\mu\al}q^\bt)\stackrel{0}{U}{\!}_{n}{\al\bt}
-(q^\chi \stackrel{0}{W}{\!}_{n'}^{\al\bt}-2q^\al
\stackrel{0}{W}{\!}_{n'}^{\bt\chi})\stackrel{0}{U}{\!}_{n}^{
\mu\chi}\stackrel{0}{U}{\!}_{n}{\al\bt}$. Here  $
\stackrel{0}{W}{\!}_{n}^{\al\bt}
=\stackrel{0}{U}{\!}_{n}^{\al\bt}-\fr12\eta^{\al\bt}\stackrel{0}{U}_{n}
$.
\par
It can be checked that the quantities $ D^\mu_{n'n} $, $
X^\mu_{n'n} $ and $ Z^\mu_{n'n} $ satisfy the conditions $
D^\mu_{n'n} u_{n\mu}=D^\mu_{n'n} \Sigma_{n\mu}=0 $, $ X^\mu_{n'n}
u_{n\mu}=X^\mu_{n'n} \Sigma_{n\mu}=0 $, $ Z^\mu_{n'n}
u_{n\mu}=Z^\mu_{n'n} \Sigma_{n\mu}=0, $ which ensure the
fulfilment of the constraint equations (\ref{con})  up to the
first order terms.
\par
Radiation arises in the second order field terms
${\stackrel{2}\phi}_{n}$ and ${\stackrel{2}B}{\!}^{\mu\nu}_n$
which are generated by the first order currents
$\stackrel{1}J_{(\phi)}$,${\stackrel{1}J}{\!}^{\,\mu\nu}_{(B)}$
 in the dilaton and axion field equations
((\ref{fep}),(\ref{feB})). These currents are constructed using
the first order quantities, so the resulting equations read \be
\Box \stackrel{2}{\phi}=4\pi\stackrel{1}J_{(\phi)},\qquad \Box
\stackrel{2}{B}{\!}^{\mu\nu}=4\pi{\stackrel{1}J}{\!}^{\,\mu\nu}_{(B)},\ee
where
 \ba \fl \stackrel{1}J_{(\phi)} =\sum_{n\neq n'} \Big\{
\fr{\al\mu}{8\pi}\int  d^2\si
\left[\left(\stackrel{0}{\dot{x}}{\!}_n ^{(\mu
}{\stackrel{1}{\dot{x}}}{\!}_n^{\nu )}-{\stackrel{0}{x'}}{\!}_n
^{(\mu } {\stackrel{1}{x'}}{}_n^{\nu )
}\right)\eta_{\mu\nu}-\fr12\stackrel{0}{U}{\!}_{n}{\stackrel{1}{x}}{\!}_n^\la
\partial_\la\right]\!\delta^{4}
(x- \stackrel{0}{x}{\!}_n (\sigma_n))\non\\\qquad\fl+
\fr{\al^2\mu}{8\pi}\int d^2\si
\stackrel{0}{U}{\!}_{n}\stackrel{1}{\phi}{\!}_{n'} +
\fr{\al\mu}{16\pi}\int d^2\si
\stackrel{0}{U}{\!}_{n}^{\mu\nu}\stackrel{1}{h}_{n' \,\mu\nu
}-\fr1{4\pi}\partial^\mu((\partial^\nu
\stackrel{1}{\phi}{\!}_n)\stackrel{1}{\psi}{\!}_{n'\,\mu\nu})+
\fr{\al}{24\pi}H^2_n \Big\},\non\ea\ba\fl
{\stackrel{1}J}{\!}^{\,\mu\nu}_{(B)} =\sum_{n\neq n'} \Big\{f
\int\limits d^2 \sigma \left[ \left(
\stackrel{0}{\dot{x}}{\!}_{n}^{[\mu}
{\stackrel{1}{x'}{\!}^{\nu]}_n} +
\stackrel{1}{\dot{x}}{\!}_{n}^{[\mu}
{\stackrel{0}{x'}{\!}^{\nu]}_n}\right) -
\fr12\stackrel{0}{V}{\!}_{n}^{\mu \nu}
\,\stackrel{1}{x}{\!}_{n}^\lambda\,\partial_\lambda\right]
\delta^{4} (x- \stackrel{0}{x}{\!}_n (\sigma_n))\non\\+
\fr{1}{8\pi} \Box
\stackrel{1}{B}{\!}^{\mu\nu}_n\stackrel{1}{h}{\!}_{n'}+
\fr{1}{8\pi}\stackrel{1}{H}{\!}_{n}^{\la\mu\nu}\partial_\la\stackrel{1}{h}{\!}_{n'}
-\fr{\al}{\pi} \Box
\stackrel{1}{B}{\!}^{\mu\nu}_n\stackrel{1}{\phi}{\!}_{n'}-\fr{\al}
{\pi}\stackrel{1}{H}{\!}_{n}^{\la\mu\nu}\partial_\la\stackrel{1}{\phi}{\!}_{n'}
\Big\}.\lb{JJ}\ea Here  brackets $(),\,[]$ denote symmetrization
and alternation over indices with the factor $1/2$, $
\psi_{\mu\nu}=h_{\mu\nu}-\fr12\eta_{\mu\nu}h$ and the D'Alembert
operator is $\Box=-\eta^{\mu\nu}\pa_\mu\pa_\nu $. The right hand
sides of the field equations contain  the first order field
quantities \ba \fl \lb{phi1}
\stackrel{1}{\phi}=\!\fr{\al\mu}{8\pi^2} \int
\fr{\e^{iq_\la\stackrel{0}x{\!}^\la_{n}}\delta(q
u_{n})\delta(q\Sigma_{n})}{q^2+2i\ep q^0}d^4 q,\qquad
\stackrel{1}B{\!}^{\mu\nu}_n=\!\fr{f}{2\pi}
\int\frac{\e^{iq_\la\stackrel{0}x{\!}^\la_{n}}
\stackrel{0}V{\!}^{\mu\nu}_n \delta(q u_n)\delta(q \Sigma_n
)}{q^2+2i\epsilon q^0}\, d^4q \non\ea and \be \lb{h1}
\stackrel{1}{h}{\!}_{\mu\nu}=\fr{4\mu G}{\pi} \int \fr{ W_{\mu\nu}
e^{-iq_\la \stackrel{0}{x}{\!}^\la _n}\delta(q u_n)\delta(q
\Sigma_n )}{q^2+2i\ep q^0}d^4 q,\lb{h1}\ee here $
\stackrel{0}V{\!}^{\mu\nu}_n=u^\mu_n\Sigma^\nu_n-u^\nu_n\Sigma^\mu_n
 $.\\
  Note that gravitational radiation in four dimensions  is
absent \cite{GaGrLe93}, so we do not consider the second order
graviton equation.  The dilaton and axion radiation power can be
computed as the reaction work given by the half sum of the
retarded and advanced fields upon the sources \cite{GMK04} and can
be presented in the form: \ba \fl P^{\mu}_{(\phi)}  =
\fr{16}{\pi}\int k^\mu
\fr{k^0}{|k^0|}|{\stackrel{1}J}_{(\phi)}(k)|^2 \delta (k^2)d^4k,
 \qquad P^{(B)\mu }  =   \fr1{\pi}\int k^\mu
\fr{k^0}{|k^0|}|{\stackrel{1}J}{\!}^{\al\beta}_{(B)}(k)|^2 \delta
(k^2)d^4k.\lb{Pp}\ea The final formula for the dilaton and axion
bremsstrahlung from the collision of two global strings can be
obtained analytically in the case of the
ultrarelativistic collision with the Lorentz factor $\gamma=(1-v^{2}%
)^{-1/2}\gg1$. We assume the BPS condition for  the coupling
constants \cite{BuDa98} $\al\mu=2\sqrt{2}\pi f$. The main
contribution to radiation turns out to come from the graviton
exchange terms. The spectrum has an infrared divergence due to the
logarithmic dependence of the string interaction potential on
distance, so a cutoff length $\Delta$ has to be introduced: \be
\fl P^{(\phi)}=\fr{200}{3}\pi G^2\al^2\mu^4
L\ka^5(f(y)+\fr1{25}f_1(y)),\qquad P^{(B)}=\fr{16\pi^3 G^2\mu^2L
f^2\kappa^5}{3 }(f(y)-f_2(y)),\lb{Totp1}\ee where $L$-length of
the string, $y=\frac{d}{\gamma\kappa\Delta}$,
$\kappa=\gamma\cos\alpha $, $\alpha$ is the strings inclination
angle and \be\fl f(y)=12\sqrt{\fr{y}{\pi}}\,
{_2F_2}\left(\fr12,\fr12;\fr32,\fr32;-y\right)
-3\ln\left(4y\e^{C}\right),\ee\be\fl f_1(y)=(1-{\rm
erf}(\sqrt{y}))\left(\fr83y^3-30y^2+114y+\fr{169}{2}\right)-\fr{\e^{-y}\sqrt{y}}{\pi}\left(
\fr{8}3y^2-\fr{94}{3}y+131 \right), \ee \be\fl f_2(y)=(1-{\rm
erf}(\sqrt{y}))\left(\fr83y^3+6y^2-6y-\fr{5}{2}\right)-\fr{\e^{-y}\sqrt{y}}{\pi}\left(
\fr83y^2+\fr{14}{3}y-7 \right), \ee $F$ is the generalized
hypergeometric function  and $C$ is the Euler's constant.
\section{Cosmological estimate}
The evolution of cosmic superstring networks was recently
discussed in Refs.(~\cite{Sa04}-\cite{Ty05}). It  was shown that
cosmic  superstrings share a number of properties of usual cosmic
strings, but there are also differences which may lead to
observational signatures. In particular, for usual cosmic strings
the probability of the loop formation $\cP$ is  of the order of
unity, whereas for F-strings $ \cP $ is $10^{-3} \le \cP \le $1
and D-strings $10^{-1} \le \cP \le $1. The cosmic superstring
network has a scaling solution and the characteristic scale  is
proportional to the square root of the reconnection probability.
The typical separation between two long strings is comparable to
the horizon size, $ \zeta (t) \simeq \sqrt{\cP} t $. The numerical
results show that the network of long strings will reach an energy
density \be\lb{ed} \rho_s = \fr\mu{{\cP} t^2}. \ee Consider the
scattering of an ensemble of randomly oriented straight strings on
a selected target string in the rest frame of the latter. Since
the dependence of the string bremsstrahlung on the inclination
angle $\al$ is smooth, we can use for a rough estimate the
particular result obtained  for the parallel strings ($\al=0$)
introducing an effective fraction $\nu$ of ``almost'' parallel
strings (roughly 1/3).  For $N$ strings in the normalization cube
$V=L^3$, we have to integrate the radiation energy released in the
collision with the impact parameter $d=x$ over the plane
perpendicular to the target string with the measure $N/L^2\cdot
2\pi x dx$. To find  the radiation power per unit time we have to
divide the integrand by the impact parameter. Multiplying this
quantity by the total number of strings $N$ to get the radiation
energy released per unit time within the normalization volume, we
obtain in the axion case: \be Q_{brem}=\int_0^L
P^0\,\nu\,\fr{N}{L^2}\,\fr{N}{V}\,2\pi dx,\ee where we can use the
Eqs. (\ref{Totp1}) for $P^0$.  Taking into account that the string
number density is related to the energy density (\ref{ed}) via \be
\fr{N}{V}=\fr{\rho_s}{\mu L},\ee and  assuming for a rough
estimate $L\sim\Delta\sim t$,  we obtain \ba\fl Q^{(\phi)}_{brem}
\simeq 800\pi^2 G^2\al^2\mu^4\nu\ga^5\ln{\ga} \fr1{{\cP} t^3}
 ,\qquad  Q^{(B)}_{brem} \simeq  64\pi^4 G^2\mu^2 \nu
 f^2\ga^5\ln\ga
 \fr1{{\cP} t^3}.\lb{QQ}\ea  Note that the realistic value
of $\ga$ is of the order of unity, while our formulas were
obtained in the $\ga\gg 1$ approximation. Still we hope to give
the correct order of magnitude estimate.
\par
Now we can calculate the energy density of the bremsstrahlung
dilaton and axions for radiation dominated Universe as a function
of time. Since dilatons and axions are massless at this stage,
their energy density scales with the Hubble constant as $H^{-4}$,
so we have the equation \be \fr{d \va}{dt}=-4H\va+Q_{brem},\ee
where $H=\frac1{2t}$. From here we obtain for the energy density
of the bremsstrahlung dilaton and axions at the moment $t > t_0$ :
 \ba\fl \va^{(\phi)} \simeq 800\pi^2 G^2\al^2\mu^4\nu\ga^5\ln{\ga}\fr{\ln{(t/t_0)}}{{\cP} t^2} ,\qquad
\va^{(B)} \simeq 64\pi^4 G^2\mu^2 \nu
f^2\ga^5\ln\ga\fr{\ln{(t/t_0)}}{{\cP} t^2} ,\ea  where $t_0$ - the
initial time of the long string formation.\par
This work  was  supported in part by RFBR grant 02-04-16949.


\section*{References}
\begin{thebibliography}{10}
\bibitem{DuKu05}
 Durer R Kunz M and Sakellariadou M 2005 {\it Phys. Lett.} B {\bf 614} 125

\bibitem{AlBrEa00}
Alexander S, Brandenberger R and Easson D 2000 {\it Phys. Rev.} D
{\bf 52} 103509

\bibitem{Po04}
Polchinski J 2004 {\it Int. J. Mod. Phys.} {\bf A20} 3413

\bibitem{DaKi05}
Davis A S and Kibble T W B 2005 {\it Contep. Phys.} {\bf 46} 313

\bibitem{AbCo00}
Abou-Zeid M and Costa M S 2000 {\it Phys. Rev.} D {\bf 61} 106007

\bibitem{GaM01}
Galtsov D V  and Melkumova E Yu 2001 {\it Phys. Rev.} D {\bf 63}
064025

\bibitem {VaVi85}
Vachaspati T and Vilenkin A 1985 {\it Phys. Rev.} D {\bf 31} 3052

\bibitem {DaQu89}
Dabholkar A J. and Quashnock J 1989 {\it Nucl. Phys.} B {\bf 333}
815

\bibitem{Da85}
Davis R L 1985 {\it Phys. Rev.} D {\bf 32} 3172; 1985 {\it Phys.
Lett.} B {\bf 180} 225

\bibitem {DaVi97}
Damour T and Vilenkin A 1997 {\it Phys. Rev. Lett.} {\bf 78} 2288

\bibitem {BaKa05}
Babichev E and Kachelrie{\ss} M 2005 {\it Phys. Lett} B {\bf 614}
1


\bibitem {ViVa85}
Vilenkin A and Vachaspati T 1985 {\it Phys. Rev.} D {\bf 35} 1138


\bibitem {GaGrLe93}
Gal'tsov D V, Grats Yu V and Letelier P S 1993 {\it Ann. of Phys.}
{\bf 224} 90

\bibitem {GMK04}
Gal'tsov D V, Melkumova E Yu and Kerner R 2004 {\it Phys. Rev.} D
{\bf 70} 045009


\bibitem {BuDa98}
Buonanno A and Damour T 1998 {\it Phys. Lett.} B {\bf 432} 51

\bibitem {Sa04}
Sakellariadou M 2005 {\it J. Cosmol. Astropart. Phys.}
JCAP0504(2005)003

\bibitem {CoMaPo05}
Copeland E Myers R C and J.~Polchinski J 2004 {\it J. High Energy
Phys.} JHEP06(2004)013

\bibitem {CoSa05}
Copeland E and Saffin P 2005 {\it J. High Energy Phys.}
JHEP0511(2005)023

\bibitem {AS05}
Avgoustidis A and Shellard E P S 2005 {\it Phys. Rev.} D {\bf 71}
123513

\bibitem {Ma04}
Martins C J A P 2004 {\it Phys. Rev.} D {\bf 70} 107302

\bibitem {Ty05}
Tye C H H, Wasserman I and Wyman M 2005 {\it Phys. Rev.} D {\bf
71} 103508
\end{thebibliography}
\end{document}